\documentclass[mathleft
]{an}
\usepackage{graphicx}
\usepackage{cite}
\usepackage{times}
\begin{document}

\Pagespan{1}{}
\Yearpublication{2007}
\Yearsubmission{2007}%
\Month{10}%
\Volume{999}%
\Issue{}%


\title{Using X-ray observations to identify the particle acceleration mechanisms in VHE SNRs and ``dark'' VHE sources}

\author{G. P{\"u}hlhofer
\thanks{Corresponding author:
  \email{Gerd.Puehlhofer@lsw.uni-heidelberg.de}\newline}
}
\titlerunning{X-ray observations of VHE SNRs and ``dark'' VHE sources}
\authorrunning{G. P{\"u}hlhofer}
\institute{
Landessternwarte, K{\"o}nigstuhl, 69117 Heidelberg, Germany}

\received{September 3, 2007}
\accepted{October 11, 2007}


\abstract{%
Very high energy (VHE) $\gamma$-ray observations have proven to be very successful in localizing Galactic acceleration sites of VHE
particles. Observations of shell-type supernova remnants have confirmed that particles are accelerated to VHE energies in supernova blast
waves; the interpretation of the $\gamma$-ray data in terms of hadronic or leptonic particle components in these objects relies nevertheless
strongly on input from X-ray observations. The largest identified Galactic VHE source class consists of pulsar wind nebulae, as detected in
X-rays. Many of the remaining VHE sources remain however unidentified until now. With X-ray observations of these enigmatic ``dark'' objects
one hopes to solve the following questions: What is the astrophysical nature of these sources? Are they predominantly electron or hadron
accelerators? And what is their contribution to the overall cosmic ray energy budget? The paper aims to provide an overview over the
identification status of the Galactic VHE source population.}

\maketitle

\section{Introduction} 

Ground-based Cherenkov telescopes detect cosmic $\gamma$-rays in the {\em Very High Energy} (VHE, 100\,GeV -
100\,TeV) domain.
In this frequency range, sources are visible
in which charged particles are accelerated to TeV energies and beyond; those particles
give rise to the detected $\gamma$-ray emission. The energetic particles can be confined inside the 
objects, close to the
actual acceleration site, like in young \linebreak%
 shell-type supernova remnants (SNRs). In other cases, the particles diffuse out into the surrounding
medium after their acceleration to high energies; in these cases, the size of the emitting region itself defines the extent of the
``source'' or object, like for example in a pulsar wind nebula (PWN). 

The truly diffuse component of the high energy cosmic ray particles is predominantly being traced in the 
MeV-GeV domain, accessible to $\gamma$-ray satellites.
Lower energy diffuse {\em electrons} are also traced through synchro\-tron emission in the radio band.
At VHE energies, localized sources dominate because of their intrinsically harder particle spectra
 compared to 
the diffuse component.

A successful VHE survey of the Galactic plane became possible with the
high sensitivity and large field of view (FoV) of the H.E.S.S.\ (High Energy Stereoscopic System) Cherenkov telescope system, 
with
$F_{\mathrm{min}}(>100\,\mathrm{GeV}) \sim 4 \times 10^{-12}\mathrm{erg\,cm^{-2}s^{-1}}$ for a $5\,\sigma$ point-source detection in 25 hrs,
and a FoV of 3$^{\circ}$ FWHM 
(Aharonian et al. 2006c).
The H.E.S.S. telescope system became operational end of 2003, and provides through its location in Namibia
an excellent view of the 
Galactic center region. 
The majority of the currently over 50 known Galactic VHE sources was discovered in the H.E.S.S. survey of the Galactic plane 
(Aharonian et al. 2005h, 2006g; Hoppe et al. 2007; Kosack et al. 2007),
or in the H.E.S.S. FoV of
observations of
other
targets in the Galactic plane 
(e.g. Aharonian et al. 2005c, 2007b).

Perhaps somewhat unexpectedly, many of the new \linebreak%
VHE-emitting sources can not readily be identified with \linebreak%
known astrophysical objects.
Therefore, Galactic VHE astronomy does not only deal with the 
identification of 
particle acceleration mechanisms and efficiencies 
in well-known astrophysical objects, 
but also with the identification of \linebreak%
sources which are so far ``only'' defined through their $\gamma$-ray emission.
Follow-up observations of those 
sources with highly sensitive X-ray instruments 
such as 
{\it XMM-Newton} provide the 
very promising possibility to identifiy those \linebreak%
enigmatic sources of high energy particles in our Galaxy.

\section{The VHE -- X-ray connection}

{\em Leptonic} 
particles can be traced through synchrotron emission in ambient magnetic fields, 
in Galactic sources most notably in the radio and X-ray domains. 
Both,
characteristic emission frequency and flux level depend on the
B-field strength. 
If the non-thermal X-ray spectrum of an object can be attributed to synchrotron emission, 
then electrons with energies of
$\sim$50\,TeV are being traced in typical interstellar magnetic fields of 3-5\,$\mu \mathrm{G}$ 
(Aharonian et al. 1997).
This electron population produces VHE $\gamma$-rays predominantly through Inverse-Compton (IC) scattering of ambient photons. 
In the TeV range, IC photon energies are of the same order as the upscattering electron energies.
The minimum 
flux level is defined through the cosmic microwave background (CMB), higher IC flux levels
can occur
if far infrared (FIR) 
or stellar photon fields
become relevant as target, either because of high surrounding fields or from the source itself.
Only in special cases like the Crab Nebula, synchrotron photons provide the dominant IC target field 
(e.g. Aharonian et al. 2004a).

In typical interstellar B-fields, the corresponding peak energy fluxes in the VHE and X-ray band are of comparable magnitude. 
The sensitivities of current X-ray detectors such as {\it XMM-Newton} are therefore ideally suited to investigate VHE IC sources.
Special care must however be taken if the VHE
particle spectrum cuts off above $\sim 10\,\mathrm{TeV}$, because 
the corresponding peak of the synchrotron flux 
shifts
into the soft
X-ray and UV band, normally not accessible because of absorption in the Galactic plane.
Moreover, as many VHE sources are extended beyond the $10^{\prime}$ scale, only a fraction of the corresponding expected synchrotron flux can be
mapped in a single X-ray telescope pointing. The B-field may not be constant throughout the VHE emitting region; in fact,
in known VHE sources it is higher near the acceleration site, e.g. close to pulsar wind termination shock,
with increased synchrotron emissivity there. 
Hence,
{\it XMM-Newton's} large FoV 
is an important feature to pin down the actual acceleration site of a VHE source. 

{\em Hadronic}
particles are visible in the $\gamma$-ray domain in nuclear collisions 
followed by $\pi^0$-decay induced $\gamma$-ray emission. The photons typically carry $\sim$20\%
of the
primary
energy 
(Kelner et al. 2006).
If such an emission region is displaced from the actual particle acceleration
site, this particle population is visible in X-rays only through synchrotron emission from {\em secondary} electrons,
close to the sensitivity limit of current X-ray detectors.
If particles are still being accelerated to high energies, then 
electrons are expected to be accelerated in conjunction with hadrons, 
with
significant {\em primary} electron X-ray synchrotron emission.
Especially in this latter case, 
the synchrotron spectrum provides a key diagnostic tool for the interpretation of the VHE emission.

Nonthermal Bremsstrahlung of energetic electrons can also be expected, both in the X-ray and the TeV energy range. In most scenarios however,
Bremsstrahlung flux levels are below synchrotron and IC levels. 
There are cases where the interpretation of hard X-ray continuum emission in terms
of either synchrotron or Bremsstrahlung is not unambiguous, 
leading to large differences in the respectively deduced electron energies, for example in Cassiopeia A \linebreak%
(e.g. Vink 2005).

\section{The Galactic VHE source catalog}

Over 50 Galactic sources of VHE $\gamma$-ray emission are
pre\-sently known. A comprehensive list of the sources 
can be found
on the web\footnote{\em
http://www.mpi-hd.mpg.de/HESS/public/HESS\_catalog.htm, http://www.mppmu.mpg.de/~rwagner/sources/, http://tevcat.uchicago.edu/}.

So far, four objects are 
identified as supernova remnant shells. RX\,J1713.7-3946 
(Aharonian et al. 2004b, 2006a, 2007a)
and RX\,J0852.0-4622
(Aharonian et al. 2005i, 2007d)
are resolved in the VHE band, and the
emission traces the respective SNR shells. RCW\,86 
(Hoppe et al. 2007a)
is also resolved, with a VHE image matching the SNR shell.
The $\gamma$-ray emission from Cassiopeia A 
(Aharonian et al. 2001, Albert et al. 2007a)
is also attributed to its shell,
though the object cannot be resolved in the VHE band.

The largest group of identified VHE sources consists of PWNe: 
The Crab Nebula 
(Weekes et al. 1989), and the H.E.S.S. detections
G0.9$+$0.1, 
MSH\,15-5{\it2}, 
Vela-X,
K3 in the Kookaburra complex,
and HESS\,J1825-137 
(Aharonian et al. 2005e, 2005g, 2006h, 2006e, 2005d).
All but the first two PWNe are spatially resolved in the VHE band. The objects are clearly identified because of an
association with a pulsar and a matching X-ray synchrotron nebula. 

PWN processes also play a role in some $\gamma$-ray binaries. 
This object class consists of point-like and variable VHE sources, where the light
curves match those from 
the corresponding
high mass X-ray binary systems.
Examples are
PSR\,B1259-63,
LS\,5039, 
and LSI\,61-303 
(Aharonian et al. 2005b, 2005f, 2006f; Albert et al. 2006b).
The nature of the compact object is not certain in the latter two objects. 

Recently, VHE emission
coincident with the
stellar cluster Westerlund 2 was detected 
(Aharonian et al. 2007e),
ope\-ning up the possibility that
VHE particle acceleration may take place in collective phenomena such as stellar winds.
The 
nature of the acceleration process is
however not yet certain.
In general,
identification in the context here means 
that the $\gamma$-ray emission 
is 
associated with a counterpart identified in lower frequency bands. 
An identification of the actual particle acceleration process and 
its
exact location inside the object is not required at this stage.

For several newly detected VHE sources,
an identification with a lower frequency band object has been suggested based on positional proximity and
energetic arguments. After some remarks on identified SNR shells and PWNe,
some associations
where one of these scenarios is likely fulfilled will be discussed. 
For few objects, an association of an extended VHE object with an X-ray binary has been suggested, 
but the situation in those cases is less clear.

\section{Shell-type supernova remnants}

All shell-type VHE SNR 
are young, and are also emitters of non-thermal X-rays. The VHE emission unambiguously
traces VHE particles in the shells, which are 
presently
being accelerated to these high energies. However, despite the fact that spectra for
the bright SNRs range from 150\,GeV to above 50\,TeV, the lever arm is not large enough to unambiguously decide whether the particles are
hadrons or leptons. The classical hadronic $E^{-2}$-type spectra can be modified at the high energy end through particle escape, and
overall through nonlinear acceleration effects, making a distinction between leptonic and
hadronic
spectra
difficult.

X-ray data of these SNRs can be used in two steps: 
Firstly, the width of thin synchrotron-emitting filaments can be used to determine the
B-field in the emission region, as the determined length scale is commonly being linked to synchrotron cooling 
(e.g. Vink \& Laming 2003, V{\"o}lk et al. 2005).
Such derived B-fields exceed the shock-compressed interstellar B-field values by far and are
typically above \linebreak%
$100\,\mu\mathrm{G}$ for young SNRs. 
Secondly, using the such derived B-fields and the total synchrotron flux, the expected
IC emission of the SNR can be calculated. While $\sim$$10\,\mu\mathrm{G}$-sized fields 
could explain the detected VHE $\gamma$-ray emission in
the VHE-emitting shells as IC radiation, in $\sim$$100\,\mu\mathrm{G}$-sized fields 
the expected IC flux is by far lower than the detected VHE flux,
unless morphological patterns like small magnetic field filling factors affecting the synchrotron emissivity 
(e.g. Lazendic et al. 2004)
would be
invoked. 
Therefore, in SNR models it is concluded that the VHE emission is of hadronic origin
(Berezhko et al. 2003, 2006, 2007).

To determine the flux of the hadronic cosmic ray particles in the source, 
the target gas density needs to be known. Thermal X-ray emission as
detected in some young SNR can provide valuable information on the shock-heated gas. 
However, for all SNRs detected in VHE emission, the environment seems strongly influenced by the particle wind of the progenitor star.
In such wind bubble setups, it has been argued that the gas heating may yet be suppressed. Moreover, 
thermal equilibrium may not have been reached, 
affecting
the determination of the emitting gas density. Special care must therefore be taken when
interpreting the gas density, in particular the limits thereof in 
RX\,J1713.7-3946 and RX\,J0852.0-4622.

\section{VHE pulsar wind nebulae}

Young pulsars like the Crab pulsar convert $\sim$50\% of their spin-down luminosity into accelerated particles. This population
is usually assumed to be dominated by leptons that were created in the Poynting-flux dominated pulsar wind. 
The particles are redistributed in energy and accelerated to high energies at a termination shock and 
radiate unpulsed synchrotron emission in the B-field also formed in the wind, as well as IC emission. 

In compact nebulae with high B-fields as the Crab Nebula ($\sim$160\,$\mu\mathrm{G}$, 
Aharonian et al. 2004a), 
the synchrotron flux strongly exceeds the IC flux, despite an additional
contribution to the IC target 
field from synchrotron photons. 
In middle-aged PWN, X-ray and VHE energy fluxes are comparable, and the
data can be used to constrain the B-field and/or the target photon fields
(FIR and stellar fields).
Examples are Vela-X and MSH\,15-5{\it2}, which exhibit a similar VHE and X-ray morphology.

In the case of HESS\,J1825-137, the 
associated
X-ray nebula G18.0-0.7 detected with {\it XMM-Newton} 
(Gaensler et al. 2003) is much
 more compact than the VHE nebula. An identification
of the two sources was nevertheless possible because of the common offset direction of both nebulae \linebreak%
 from the powering pulsar, together with
spectral imaging of the H.E.S.S. source 
(Aharonian et al. 2006b).
The different sizes can be explained in a ``relic'' PWN scenario 
by the larger lifetime of the IC-emitting VHE particles, compared to the higher-energy synchrotron-emitting (and thus cooling) 
particles 
(Aharonian et al. 1997).
The fact that the nebula is expanding asymmetrically can be explained in a ``crushed'' PWN scenario
like in Vela-X 
(Blon\-din et al. 2001)
by the interaction with an asymmetrically evolving SNR reverse shock.
In general, ISM density gradients might cause such an effect even in a non-SNR environment.
The offset of the VHE peak intensity from the pulsar ($\sim$0.3$^{\circ}$) can be related to a higher injection efficiency of the pulsar in
the past. 

Comparing the pulsar population to the H.E.S.S. Galactic plane survey, it turns out that energetic pulsars 
are indeed very likely associated with VHE emission 
(Aharonian et al. 2007c).
For some VHE sources, the probability for a chance coincidence
with an energetic pulsar is reasonably low to 
identify
the sources as {\em VHE PWN candidates}, even without the detection of a synchrotron nebula.
Examples are HESS\,J1718-385, 
HESS\,J1809-193, 
and HESS\,J1912+101 
(Aharonian et al. 2007c, Komin et al. 2007, Hoppe et al. 2007b).
For HESS\,J1809-193, {\it Chandra} data have 
mean\-while
indeed revealed the existence of an X-ray PWN around the corresponding pulsar
(Kargaltsev \& Pavlov 2007).

\section{``VHE composite'' supernova remnants}

Several H.E.S.S. survey sources are positionally coincident with radio SNRs. 
For those, a unique identification with the {\em shells} is however not
possible based on the VHE and radio data alone. 
For HESS\,J1813-178 and HESS\,J1640-465, the SNR shells G12.8-0.0 
(Brogan et al. 2005)
and G338.3-0.0 
(Whiteoak \& Green 1996)
are too compact to be resolved in the VHE band. Those remnants were
however easily investigated in single {\it XMM-Newton} pointings 
(Funk et al. 2007b, 2007a).
Surprisingly, the {\it XMM-Newton} data revealed
new PWN candidates inside the shells, whereas no X-ray emission was detected from the shells themselves. 
The PWN candidates were unexpected because no radio pulsars are known inside the remnants. 
The identification of the X-ray sources as PWN seems 
nevertheless
plausible because of their morphology and SNR location. {\it Chandra}
follow-up observations of the PWN candidate inside HESS\,J1813-178 have been performed in the mean time with the aim to confirm the PWN
nature 
(Helfand et al. 2007).

The data establish G12.8-0.0 and G338.3-0.0 as composite SNRs. Because the VHE emission cannot be 
 unambiguously identified with either
the respective SNR shell or PWN yet, 
one might label them ``VHE composites'' for now.

Also HESS\,J1804-216 can be similarly classified. In this case however,
the radio SNR (W30, 
Kassim \& Weiler 1990)
is much more extended
than the VHE source, which might morphologically be associated with part of the (not well defined) shell or with the SNR interior.
The X-ray PWN around PSR\,B1800-21 as detected with {\it Chandra} represents a plausible counterpart
(Kargaltsev et al. 2007a, 2007b). \linebreak%
However, as the proper motion of the pulsar
(Brisken et al. 2006)
disfavours its association with W30, other PWN (without radio pulsars)
could be present inside W30. 
Indeed, a {\it Suzaku} pointing on
the center of the H.E.S.S. source has revealed another possible, albeit yet unidentified X-ray counterpart
(Bamba et al. 2007).
Nonetheless, 
{\it XMM-New\-ton} 
data might be
required to achieve sufficient coverage of the VHE source to identify all possible counterparts, and to examine the faint
extended PWN around PSR\,B1800-21 (or other PWN candidates) to allow for a morphological association between the X-ray and VHE source.
Also, the connection of the VHE particles to the Northern SNR shell,
which based on ROSAT data was identified as a thermal X-ray emitter
(Finley \& Oegelman 1994),
is an attractive target for {\it XMM-Newton} investigations.

\section{``Dark'' VHE sources}

Several VHE sources entirely lack plausible counterparts in lower energy bands, and have therefore been called ``dark'' VHE sources
(Aharonian et al. 2005h, 2006g; Kosack et al. 2007). 
Physically, a ``dark'' VHE emitter might either be purely driven by hadronic
emission (though some level of synchrotron emission is
inevitably expected from secondary electrons if the $\gamma$-rays are produced in hadronic collisions), 
or by exotic processes (such as dark matter emission).

Matsumoto et al. (2007)
argue that the X-ray limit
achie\-ved in a deep {\it Suzaku} exposure of HESS\,J1616-508 excludes a leptonic scenario for any plausible B-field,
and call the VHE source ``dark''. 
On the other hand,
the energetic pulsar PSR\,J1617-5055 represents a 
possible
counterpart in an offset PWN scenario 
(Aharonian et al. 2006g, Landi et al. 2007).
Here, the expected X-ray synchrotron emission 
depends on the extrapolation from the UV 
band
that corresponds
to the VHE IC flux, and the limited {\it Suzaku} FoV might not have covered the entire 
VHE-emiting region.
Even PSR\,J1614-5048 might power (part of) the
source,
though this scenario is
challenging because of the high required conversion efficiency of spin-down to VHE luminosy of \linebreak%
$\sim$10\%, and the large angular
offset.
Further investigations are hence required to clarify the nature of HESS\,J1616-508.

It was argued by 
Yamazaki et al. (2006)
that the absence of significant X-ray emission
may be a signature for old 
but still VHE emitting SNRs. 
Significant VHE 
emission from hadronic particles could still be expected, 
either from the shell or from particles encountering nearby molecular clouds, whereas primary 
electrons 
are not any more energetic enough to emit X-ray synchrotron radiation. 
The X-ray emission may therefore be dominated by secondary electrons,
with TeV to X-ray flux ratios (expressed as \linebreak%
$R_{\mathrm{TeV/X}} = F_{\gamma, 1-10\,\mathrm{TeV}} / F_{\mathrm{X},
2-10\,\mathrm{keV}}$) of 
$\sim$100 or more, compared to young SNRs with values of less than $\sim$2. 

Nonetheless, other object types may also exhibit such large TeV to X-ray flux ratios, 
therefore further broadband data are 
needed in each case to establish the nature of 
individual VHE sources.

\section{Possible interaction with molecular clouds}

Molecular clouds 
merely
serve 
as targets of energetic particles, whereas the actual acceleration site may be
disconnected from the emitting cloud. 
A prominent example
for such a scenario is the diffuse VHE emission 
from the Galactic ridge 
near 
the Galactic center 
(Aharonian et al. 2006d).

A good signature for such a scenario is the close spatial correlation between VHE and CO/CS emission tracing molecular gas.
VHE emission detected with H.E.S.S. close to SNR W28,
which 
is indeed coincident with CO emission located in a
distance similar to 
the SNR, might be caused by
particles that were 
accelerated in W28 
(Rowell et al. 2007).
Similar scenarios could explain the emission from MAGIC\,J0616+225 inside IC\,443 
(Albert et al. 2007c),
and HESS\,J1834-087 coincident with the interior of the SNR
W41
(Albert et al. 2006a).
The faint flux level possibly detected with {\it XMM-Newton} from HESS\,J1834-087 is compatible with
secondary electron emission 
(Tian et al. 2007).

\section{Prospects for XMM-Newton observations}

The high sensitivity and large FoV of {\it XMM-Newton} make this instrument ideally suited for the investigation of VHE $\gamma$-ray sources. 
The large FoV is required to identify previously unknown objects such as the PWN candidates described earlier.
Jointly with VHE detectors, {\it XMM-Newton} will thus probe the population of radio-dim pulsars.
{\it Chandra} can effectively be 
used to detect compact PWN; however, those objects are likely influenced by the pulsar geometry or motion and do
not reflect the large-scale propagation of electrons that give rise to the VHE emission.
{\it XMM-Newton's} sensitivity is required to trace those 
extended 
and offset nebulae in low B-field environments.
Also the number of VHE-emitting SNR shells is expected to grow, and the FoV and sensitivity of {\it XMM-Newton} will be needed to pin down the nature
of the VHE particles, both through primary and secondary electron emission. 
The discovery potential for new sources or even new source types in case of ``dark'' VHE source observations is large.

Together with the new VHE instruments in the Northern Hemisphere that have recently come online (MAGIC, VERITAS), 
with H.E.S.S. and its extension 
to higher sensitivity and lower energy threshold (H.E.S.S. phase II), 
and with the next generation of instruments such as CTA\footnote{\em http://www.mpi-hd.mpg.de/CTA}
that could start operation early in the next decade, 
the continuous availability of {\it XMM-Newton} will provide indispensable input for the understanding
of the high energy particle accelerators in our Galaxy.

\acknowledgements
G.P. acknowledges support by the German \linebreak
Ministry for Education and Research (BMBF) through DESY \linebreak
grant  05\,CH5\,PC1/6 and  DLR grant 50\,OR\,0502.



\end{document}